\documentclass[aps,12pt,preprintnumbers,
                secnumarabic,nofootinbib]{revtex4}

\usepackage{epsfig}
\usepackage{bbm}
\usepackage{latexsym}

\def\ri{{\rm i}}
\def\re{{\rm e}}
\def\ve{\varepsilon}
\def\ps{{\partial \!\!\!/}}

\begin{document} 

\title{\Large Ultraviolet Regularisation in de Sitter Space}

\author{Bj\"orn~Garbrecht}
\affiliation{School of Physics \& Astronomy, The University of Manchester,
Oxford Road, Manchester M13 9PL, United Kingdom}

\preprint{MAN/HEP/2006/14}

\begin{abstract}
The ultraviolet regularisation of Yukawa theory in de Sitter space is
considered. We rederive the one-loop effective Candelas-Raine potentials,
such that they agree with the corresponding Coleman-Weinberg potentials
in flat space.
Within supersymmetry, this provides a mechanism for the lifting
of flat directions during inflation. For the purpose of calculating
loop integrals, 
we employ the dimensional regularisation
procedure by Onemli and Woodard and
show explicitly that the resulting self-energies are also invariant.
This implies the absence anomalous de Sitter breaking terms, which are
reported in the literature.
Furthermore, transplanckian effects do not necessarily leave an imprint
on the spectrum of cosmic perturbations generated during inflation.

\end{abstract}


\maketitle

\section{Introduction}

Quantum theory in de Sitter space predicts a scale-invariant
primordial spectrum of density fluctuations generated during
inflation~\cite{ChibisovMukhanov:1981,Starobinsky:1982,Hawking:1982,GuthPi:1982,BardeenSteinhardtTurner:1983}.
This effect is in compelling accordance with observation
(see {\it e.g.}~\cite{Spergel:2006}) and it can be calculated by solving the
field equations of motion in a de Sitter background at tree level. There
is now strong evidence for a small deviation from
scale-invariance~\cite{Spergel:2006}, this can however be explained by the
fact that in all realistic inflationary models the background is not
exactly de Sitter space, leaving the methods for computing density
perturbations still valid.

As an example for a loop calculation, one can compute the
Coleman-Weinberg~\cite{ColemanWeinberg:1973} effective potential in curved
space-times using methods devised by DeWitt~\cite{DeWitt:1975}. In
the setting of a de Sitter background, these effective potentials have been
calculated for fermion and scalar loops by Candelas and
Raine~\cite{CandelasRaine:1975}. Their results have recently been rederived for
the case of Yukawa theory~\cite{MiaoWoodard:2006}
and have lead to the proposal of a
new relaxation mechanism for the cosmological constant~\cite{Prokopec:2006}.
In section~\ref{section:Veff}, we rederive the Candelas-Raine potentials
and confirm the original results~\cite{CandelasRaine:1975}.
As a consistency check, we confirm that the results presented here have the
advantage of reducing to the flat-space limit~\cite{ColemanWeinberg:1973}
when taking the expansion rate to zero. We point out that the
Hubble induced mass terms arising from the effective potential
can be of relevance for supersymmetric models of inflation.

Loops in de Sitter space have also been computed to obtain
self-energies~\cite{ProkopecTornkvistWoodard:2002,ProkopecTornkvistWoodard:2002:2,ProkopecWoodard:2003:photon1,ProkopecWoodard:2002:photon2,ProkopecPuchwein:2004,ProkopecPuchwein:2005,ProkopecWoodard:2003,GarbrechtProkopec:2006},
which induce corrections to the field equations of motion.
For this type of calculation, a powerful ultraviolet regularisation procedure
has been introduced by Onemli and
Woodard~\cite{OnemliWoodard:2002}.
Employing this technique, it is reported that local de Sitter breaking terms
occur, which have been interpreted as anomalous so far~\cite{ProkopecWoodard:2003,ProkopecTornkvistWoodard:2002,ProkopecTornkvistWoodard:2002:2,ProkopecWoodard:2002:photon2,GarbrechtProkopec:2006}.
An ultraviolet induced breaking of the de Sitter symmetry also plays
a role in context of the transplanckian
problem~\cite{BrandenbergerMartin:2000,BrandenbergerMartin:2000:2,Kempf:2000},
where time translation and boost invariance is assumed to be explicitly broken,
such that one may suspect these effects to be related.
The answer is that the boundary conditions
imposed to effectively take account of transplanckian effects decouple
from renormalisations in four dimensional de Sitter
space~\cite{CollinsHolman:2005,CollinsHolman:2005:2}.
Nonetheless, it is an interesting question whether one can explicitly
construct an invariant regularisation procedure in field theory or whether
ultraviolet effects necessarily lead to the breakdown of de Sitter invariance.
In section~\ref{section:self-energies}, it is shown that
the procedure by Onemli and Woodard in fact preserves invariance,
and as an example, this method is applied to Yukawa
theory. The de Sitter breaking terms are shown to be not anomalous but cancel
with contributions which have been negelected so far.

\section{Propagators \& Effective Potentials}
\label{section:Veff}

We discuss Yukawa theory as described by the Lagrangian
\begin{equation}
\label{Lagrangian}
{\cal L}=
\sqrt{-g}\left\{
-\frac 12 g^{\mu\nu}(\partial_\mu \phi)(\partial_\nu \phi)
-\frac 12 m_\phi^2 \phi^2
+\bar\psi \nabla \!\!\!\!/ \psi
-m \bar \psi \psi
-\mu^{2-\frac D2} f \psi \bar \psi \psi
\right\}
\,,
\end{equation}
where $\mu$ is a constant of mass dimension, which is introduced to ensure that
the Yukawa-coupling $f$ is dimensionless for any space-time dimension $D$,
and $\nabla\!\!\!\!/$ denotes the covariant derivative acting on spinors.
We choose conformal coordinates for de Sitter space expanding
at the Hubble rate $H$, such that
\begin{equation}
g_{\mu\nu}=a^2 \eta_{\mu\nu}\,,
\end{equation}
where $a=-1/(H\eta)$ is the scale factor, $\eta \in ]-\infty,0[$ is
the conformal time and
\begin{equation}
\eta_{\mu\nu}={\rm diag}(-1,\!\!\underbrace{1,...,1}_{D-1\; {\rm times}}\!)
\nonumber
\end{equation}
is the $D$-dimensional Minkowski metric.
When expressed in these coordinates, the Lagrangian~(\ref{Lagrangian})
takes the form~\cite{ProkopecWoodard:2003}
\begin{equation}
\label{Lagrangian:deSitter}
{\cal L}=
-\frac 12 a^{D-2} \eta^{\mu\nu}(\partial_\mu \phi)(\partial_\nu \phi)
-\frac 12 a^D m_\phi^2 \phi^2
+\left( a^{\frac{D-1}{2}} \psi \right) \ri \ps \left(a^{\frac{D-1}{2}} \psi\right)
-a^D m \bar \psi \psi
-a^D \mu^{2-\frac D2} f \phi \bar \psi \psi
\,.
\end{equation}
Here, we assume $\langle \phi \rangle=0$ and regard the effective potentials
as functions of $m$ and $m_\phi$, respectively. The reason is purely
notational, since alternatively, we could set $m=0$ and take
$\mu^{2-\frac D2} f \langle \phi \rangle$ with $\langle \phi \rangle\not=0$
as the mass term. Likewise, the scalar mass $m_\phi$ could be substituted
by introducing a self-interaction for $\phi$ and redefining the vacuum
expectation value such that $\langle \phi \rangle\not=0$.

The separation between two coordinate points is ($\eta=x^0$)
\begin{equation}
\Delta x^2(x;x^\prime)= -(|\eta-\eta^\prime| -\ri \epsilon)^2
+\sum_{i=1}^D|x^i-{x^\prime}^i|^2
\,,
\end{equation}
where the $\ri \epsilon$-term is introduced according to the
Feynman-pole prescription (see \emph{e.g.}~\cite{ProkopecPuchwein:2004}).
This is however not the same as the physical
geodesic distance  $\ell(x;x^\prime)$ between two points.
The function $\ell(x;x^\prime)$ is de Sitter invariant and most conveniently
expressed in terms of the also de Sitter invariant function $y(x;x^\prime)$
as
\begin{eqnarray}
 y &=& 4\sin^2\Big(\frac12 H\ell\Big) = aa^\prime H^2 \Delta x^2
\label{y}\,,
\end{eqnarray}
where here and in the following we abbreviate $a=a(\eta)$ and
$a^\prime=a(\eta^\prime)$.

The concept of point-splitting
regularisation~\cite{DaviesFullingUnruh:1976,Christensen:1976,Christensen:1978}
is to expand two-point functions in a series
\begin{equation}
\cdots + a_{-1}\frac 1{\ell^2} + c + a_0 \log \ell + a_1 \ell^2 +\cdots\,,
\end{equation}
where the $a_i$ and $c$ are constants. Each term is manifestly covariant,
in particular, one can subtract the terms which are ultraviolet divergent
when $\ell \rightarrow 0$, without breaking general coordinate invariance.
Namely, these divergent contributions are the negative powers of $\ell$
and the $\log \ell$ term.
When noting that
\begin{equation}
\label{y:expanded}
y=(H\ell)^2 + O\left([H\ell]^6\right)
\,,
\end{equation}
we can replace $y\rightarrow (H\ell)^2$ in these expansions, as long as the
expansion does only range over the coefficients
$a_{-1}$, $a_0$ and $c$ and and as long $\ell\alt H^{-1}$
is of subhorizon scale. This is the regularisation procedure we follow in this
section.

Let us first construct the one-loop effective potential for fermions
in de Sitter background~\cite{CandelasRaine:1975}. 
The Feynman function is given by
\begin{equation}
\ri S(x,x^\prime)=
\frac
{
\int {\cal D}\psi{\cal D}  \bar \psi\,
\psi(x) \bar \psi(x^\prime)
\re^{\ri \int d^D  x {\cal L}}
}
{
\int {\cal D}\psi{\cal D}  \bar \psi \,
\re^{\ri \int d^D  x {\cal L}}
}\,,
\end{equation}
and it satisfies
\begin{equation}
\label{Eq:GreenF:fermion}
\sqrt{-g}(\ri \nabla\!\!\!\!/-m)\ri S(x,x^\prime)= \ri \delta^D (x-x^\prime)
\,.
\end{equation}
This equation may be solved in de Sitter background using the
ansatz~\cite{MiaoWoodard:2006,Prokopec:2006}
\begin{equation}
{\rm i} S(x;x^\prime)=a\left(
a^{-\frac{D+1}{2}}{\rm i}\partial\!\!\!/ a^{\frac{D-1}{2}}+ m
\right)
(aa^\prime)^{-\frac 12}
\sum_{\pm}{\rm i} S_\pm(y) \frac{1\pm \gamma^0}{2}
\,,
\end{equation}
such that we find, when substituting above into
Eqn.~(\ref{Eq:GreenF:fermion}),
\begin{equation}
\left[
(y^2-4y)\frac{\partial^2}{\partial y^2}-D(2-y) \frac{\partial}{\partial y}
+\frac D2 \left(\frac D2-1\right)-\ri \gamma^0 \frac mH +\frac{m^2}{H^2}
\right]
\sum_\pm S_\pm(y)\frac{1\pm \gamma^0}{2}=0\,.
\end{equation}
The solution to the latter equation is a hypergeometric
function~\cite{CandelasRaine:1975,MiaoWoodard:2006,Prokopec:2006},
\begin{equation}
\label{Spm}
\ri S_{\pm}=\frac{H^{D-2}}{(4 \pi)^{\frac D2}}
\frac{
\Gamma\left(\frac D2 -1 \mp  \ri \frac mH\right)
\Gamma\left(\frac D2  \pm  \ri \frac mH\right)
}
{
\Gamma\left(\frac D2\right)
}
\,
{}_2F_1\left(
\frac D2 -1 \mp \ri \frac mH,
\frac D2 \pm \ri \frac mH;\frac D2;1-\frac y4
\right)
\label{Prop:Fermion}
\,.
\end{equation}

Note that for the purpose of point-splitting regularisation, we could
set $D=4$ from the outset.
In the following expansion however, we keep the leading terms in general
space-time dimension, as they are needed for the dimensional
regularisations taken out in section~\ref{section:self-energies}.
For the subleading
(logarithmic and constant) terms, we readily take $D=4$ ($\ve=0$).
Having said this, we expand two-point function $S_\pm$ in $y$ and find,
when and writing $D=4-\ve$,
\begin{eqnarray}
\label{Spm:Expanded}
\ri S_{\pm}&=&
\frac{H^{2-\varepsilon}}{4\pi^{2 -\frac{\ve}{2}}}
\Gamma\left(1-\frac\ve2\right)\frac{1}{y^{1-\frac{\ve}{2}}}
\\
&&+\frac{\mp \ri Hm + m^2}{16\pi^2}
\left[
\log\frac y4+\psi\left(1 \mp \ri \frac mH\right)
+\psi\left(2 \pm \ri \frac mH\right)
-1 + 2\gamma_{\rm E}
\right]
\nonumber
\,.
\end{eqnarray}
Next, we insert this into the fermion propagator~(\ref{Prop:Fermion})
and consistently keep only terms which are $\sim y^{-1}$, $\sim \log y$
and constant in $D=4$, such that we obtain
\begin{eqnarray}
\label{Prop:Fermion:Expanded}
\ri S(x,x^\prime) &=&
(aa^\prime)^{-\frac{3-\ve}{2}} \frac{1}{4\pi^{2-\frac{\ve}{2}}}
\Gamma\left(1-\frac{\ve}{2}\right)
\ri \ps\frac{1}{(\Delta x^2)^{1-\frac{\ve}{2}}}
+(aa^\prime)^{\frac{\ve}{2}}\frac{m}{4\pi^{2-\frac{\ve}{2}}}
\Gamma\left(1-\frac{\ve}{2}\right)
\frac{1}{(\Delta x^2)^{1-\frac{\ve}{2}}}
\nonumber
\\
&&
+\frac{1}{16\pi^2}
\left(
m^3 -2\ri \gamma^0 H m^2 + H^2 m
\right)
\\
&&\times
\left(
-1+2\gamma_{\rm E} + \log \left(aa^\prime H^2 \Delta x^2 \right) - 2\log 2
+\psi\left(1 - \ri \frac mH\right)+\psi\left(1 + \ri \frac mH\right)
\right)
\nonumber
\,.
\end{eqnarray}

The one-loop effective potential can now be easily calculated
using~\cite{DeWitt:1975,CandelasRaine:1975}
\begin{equation}
\frac{\partial V_{\rm eff}}{\partial m}=-{\rm tr} \sqrt{-g}\, \ri S(x,x)
\,.
\end{equation}
Noting that in the limit $m\gg H$, the Euler functions have the asymptotic
property
\begin{equation}
\psi\left(1 - \ri \frac mH\right)+\psi\left(1 + \ri \frac mH\right)
\sim \log \frac {m^2}{H^2}
\,,
\end{equation}
we obtain~\cite{CandelasRaine:1975,others,MiaoWoodard:2006}
\begin{eqnarray}
\label{Veff:fermion}
V_{\rm eff}^\psi&=&-\frac{m^2}{2\pi^2}\frac{1}{\varrho^2}
+\frac{1}{16\pi^2}
\Bigg\{
-m^4 \log(\varrho^2 m^2) -2 H^2 m^2 \log(\varrho^2 m^2)
\\
&&
+\left(\frac 32 -2 \gamma_{\rm E} +\frac 12 \log 2\right) m^4
+\left(4 - 4 \gamma_{\rm E} + \log 2 \right)H^2 m^2
\Bigg\}
\,.
\nonumber
\end{eqnarray}
Out of the terms in curly brackets,
important are only the logarithmic ones, since  the analytic contributions
are regularisation scheme-dependent and can always be canceled by adding 
counterterms to the Lagrangian. In addition, we have introduced a constant
physical cutoff scale $\varrho$ with the dimension
of a length, which is used to regulate expressions which are divergent as
$aa^\prime \Delta x^2$ goes to zero. An important consistency check is to
note that in the limit $H\rightarrow 0$, the above expression reduces to the
celebrated Coleman Weinberg potential~\cite{ColemanWeinberg:1973}.
The same result has been derived by Miao and Woodard~\cite{MiaoWoodard:2006},
who suggest to use $\varrho=H^{-1}$ as a regulator. This is motivated by
imposing the renormalisation condition that the scalar field remains free and
massive in the large $H$ limit. In turn, the scalar mass and couplings then
diverge logarithmically when $H\to 0$. Such a behaviour may lead to a solotion
to the cosmological constant problem~\cite{Prokopec:2006}.
The numerical factors in the terms presented here and in~\cite{MiaoWoodard:2006,others}  differ from the
earlier results~\cite{CandelasRaine:1975}. Note however, that the latter
effective potentials for the fermion and  for the scalar loop do not
reduce to the Coleman-Weinberg form as $H \rightarrow 0$, albeit
being well defined in that limit. In particular, all contributions
$\sim m^4 \log m$ and $\sim m_\phi^4 \log m_\phi$ cancel,
in disagreement with the flat-space result\footnote{
When comparing Eqn.~(\ref{Veff:fermion})
of this article with the corresponding Eqn.~(30) of
Ref.~\cite{CandelasRaine:1975} (Eqn.~(\ref{Veff:scalar}) here is to be
compared with Eqn.~(21) of Ref.~\cite{CandelasRaine:1975} for the scalar
case), one should be aware that the integrals over the Euler-$\psi$ functions
in Ref.~\cite{CandelasRaine:1975} comprise logarithms at leading order in
$H/m$.
}.
Our regularisation procedure differs, but we point out that this is
not the origin of the disagreement,
which should be attributed to a
calculational mishap or a typo in Ref.~\cite{CandelasRaine:1975}.
A result very similar
to Eqn.~(\ref{Veff:fermion}) is reported by Elizalde and
Odintsov~\cite{ElizaldeOdintsov:1995}.
The difference lies within renormalisation-scheme dependent terms, but
moreover an overall factor of minus one occurs when comparing the
coefficients of the logarithms.

The scalar case is more familiar, we therefore go into less details.
The Green function for the scalar field is of
the following de Sitter invariant
form~\cite{ChernikovTagirov:1968,OnemliWoodard:2002,ProkopecPuchwein:2004,GarbrechtProkopec:2006}, 
\begin{equation}
    {\rm i}\Delta(x,x')=\frac{\Gamma(\frac{D-1}{2}+\nu)
    \Gamma(\frac{D-1}{2}-\nu)}{(4\pi)^{\frac{D}{2}}
    \Gamma(\frac{D}{2})}H^{D-2}\;
{}_2F_1\Big(\frac{D-1}{2}+\nu,\frac{D-1}{2}-\nu,\frac{D}{2},1-\frac{y}{4}\Big)
\label{Prop:Scalar}
\,,
\end{equation}
where
\begin{equation}
\nu=\left[\left(\frac{D-1}{2}\right)^2-\frac{m_\phi^2+\xi R}{H^2}
      \right]^\frac 12
\,,
\label{nuD}
\end{equation}
$R = D(D-1)H^2$ denotes the Ricci scalar curvature of de Sitter space,
and $\xi$ is the coupling constant of the scalar field to
curvature
($\xi=(D-2)/(4D-4)$ corresponds to conformal coupling,
$\xi=0$ to minimal coupling).

The expanded version of the scalar propagator corresponding to the fermionic
case~(\ref{Prop:Fermion:Expanded}) is found to be~\cite{ProkopecPuchwein:2004,GarbrechtProkopec:2006}
\begin{eqnarray}
\ri \Delta(x;x^\prime)&=&
\frac{H^{2-\varepsilon}}{4\pi^{\frac D2}}
\Gamma\left(\frac D2-1\right)\frac{1}{y^{\frac D2-1}}
\\
&&+\frac{H^2}{16\pi^2}\left(\frac{m_\phi^2}{H^2}-2\right)
\left[
\log\frac y4 + \psi\left(\frac 32 +\nu\right)+ \psi\left(\frac 32 -\nu\right)
-1+ 2\gamma_{\rm E}
\right]
\nonumber\,.
\end{eqnarray}
From this, we now derive the scalar potential
using~\cite{DeWitt:1975,CandelasRaine:1975}
\begin{equation}
\frac{\partial V_{\rm eff}}{\partial m_\phi^2}={\frac 12} \sqrt{-g}\, \ri \Delta(x,x)
\,,
\end{equation}
such that we obtain for $m_\phi \gg H$~\cite{others}
\begin{eqnarray}
\label{Veff:scalar}
V_{\rm eff}^\phi&=& \frac{m_\phi^2}{8\pi^2 \varrho^2}
+\frac{1}{16 \pi^2}
\Bigg\{
\frac 14 m_\phi^4 \log\left(\varrho^2 m_\phi^2\right)
- H^2 m_\phi^2 \log\left(\varrho^2 m_\phi^2\right)
\\
&&
+\left(\frac 38 + \frac 12 \gamma_{\rm E} - \frac 18 \log 2 \right)m_\phi^4
+\left(2-2 \gamma_{\rm E} +\frac 12 \log 2 \right) H^2 m_\phi^2
\Bigg\}
\,.
\nonumber
\end{eqnarray}

Taking $H \rightarrow 0$, agreement with the
Coleman-Weinberg~\cite{ColemanWeinberg:1973}
result is found also for the scalar loop.
Just as for the fermionic case, in the final expression
reported by Candelas and Raine~\cite{CandelasRaine:1975}, all logarithmic
terms cancel in the flat-space limit $H\rightarrow 0$, in disagreement
with Coleman and Weinberg~\cite{ColemanWeinberg:1973}. Again, this should
not be attributed to the approach or regularisation used
in~\cite{CandelasRaine:1975}, but to a minor calculational mistake.

Our results~(\ref{Veff:fermion}) and~(\ref{Veff:scalar}) imply in particular
for $m=m_\phi$ that
\begin{equation}
\label{Veff:SUSY}
4 V_{\rm eff}^\phi+V_{\rm eff}^\psi=
-\frac 3{8\pi^2}H^2 m^2\log\left(\varrho^2 m^2\right)
+\frac{1}{16\pi^2}\left(12-12\gamma_{\rm E} + 3\log 2\right)H^2m^2
\,,
\end{equation}
and that in Minkowski space ($H = 0$) fermionic and scalar
contributions cancel, as they should when assuming the same number of
degrees of freedom and the same mass spectrum.
De Sitter space with $H\not=0$ exhibits
however supersymmetry breaking due to the different curvature coupling
of fermions and scalars.

This finding may be of importance for inflationary models when replacing
$m$ and $m_\phi$ by $\kappa \langle S \rangle$, where $S$ denotes the
slowly rolling inflaton field and $\kappa$ its superpotential
coupling to other chiral multiplets.
In $F$-term inflation, similar terms
are expected from supergravity corrections~\cite{CopelandLiddleLythStewartWands:1994,DineRandallThomas:1995,Panagiotakopoulos:1997,Panagiotakopoulos:2004}
and have consequences for predictions of the spectrum of primordial
density fluctuations~\cite{JeannerotPostma:2005,BasteroGilKingShafi:2006}.
For a minimal K\"ahler potential within $F$-term inflation and generally
within $D$-term inflation, these kind of
corrections originating from supergravity are expected to be
absent~\cite{KoldaMarch-Russell:1998},
whereas the de Sitter background induced corrections presented here are still
there and should be taken into account. Note that
when assuming the renormalisation scale to be larger than the mass terms,
$\rho^{-1}>m$, and the derived mass-square corrections are always positive.

As a next step, it will be important to derive the one-loop potential for
gauge bosons and gauginos in the loop. Using the above results for the scalar
and fermion case for a conjecture and
taking into account the dimension of the Standard Model
gauge group as well as typical values for the gauge coupling constants
suggested by gauge coupling unification, mass corrections which are of order
of the Hubble rate might arise also during the postinflationary
eras~\cite{BuchbinderOdintsov:1985}.
This intriguing possibiltity for a
universal mechanism of lifting flat directions of the Mimimal
Supersymmetric Standard Model is subject of ongoing studies.

\section{Self-Energies}
\label{section:self-energies}

The above expressions for the effective potentials are constructed from the
coincidence limits of the fermionic and scalar Green functions, and
the occurring ultraviolet divergences are renormalised by subtracting terms
which become infinite as we take the short-distance regulator $\varrho$ to
zero. This procedure apparently bears some similarity with  a cutoff
regularisation in momentum space. One should however be aware of the fact that
a large momentum cutoff of a two-point function in momentum space does
not result in a short distance cutoff of its Fourier transform in position
space. Therefore, the simple point-splitting method employed in the
previous section is not applicable to the case where finite distance effects
are of importance.

This can be seen when studying self-energy functions. Examples are the
vacuum polarisation for scalar electrodynamics~\cite{ProkopecTornkvistWoodard:2002,ProkopecTornkvistWoodard:2002:2,ProkopecWoodard:2003:photon1,ProkopecWoodard:2002:photon2,ProkopecPuchwein:2004,ProkopecPuchwein:2005}
and the fermion self-energy~\cite{ProkopecWoodard:2003,GarbrechtProkopec:2006}
in de Sitter background. The self-energies, which are two point functions,
give rise to corrections to the free field equations, which are in general
nonlocal. Momentum space experience is telling us that we obtain from
ultraviolet divergent integrals logarithms of functions involving the external
momentum $q$ as well as the mass of the particles running in the loop.
These contributions are of utmost importance for the
predictions of a theory, and therefore a position space technique to separate
these terms from the divergent parts is needed.
A powerful method to achieve this within
dimensional regularisation has been suggested by
Onemli and Woodard~\cite{OnemliWoodard:2002}.

Some results obtained using this
technique seem to introduce as a common feature local operators~\cite{ProkopecWoodard:2003,ProkopecTornkvistWoodard:2002,ProkopecTornkvistWoodard:2002:2,ProkopecWoodard:2003:photon1,ProkopecWoodard:2002:photon2,ProkopecPuchwein:2004,ProkopecPuchwein:2005,GarbrechtProkopec:2006}, which
violate de Sitter invariance, since they are proportional to $\log a(\eta)$
or equivalently $t$, which is the comoving time related to the scale factor
as $a(\eta)=\re^{Ht}$. This observation has been interpreted as a
perturbation theory anomaly so far~\cite{ProkopecWoodard:2003,ProkopecTornkvistWoodard:2002,ProkopecTornkvistWoodard:2002:2,ProkopecWoodard:2002:photon2,GarbrechtProkopec:2006}. However, we point out here that these terms are completely
canceled by an ostensibly negligible nonlocal contribution. Therefore,
with the choice of a de Sitter invariant counterterm, the procedure of
ultraviolet regularisation does not break de Sitter invariance. We emphasise
that for the case of a minimally coupled massless scalar field or gravity
the treatment of ultraviolet divergences suggested here,
yet leaves behind de Sitter breaking terms originating from the non-invariant
propagators.

Let us introduce local counterterms by adding
\begin{equation}
\delta {\cal L}=
\delta Z_2
\left( a^{\frac{D-1}{2}} \psi \right) \ri \ps \left(a^{\frac{D-1}{2}} \psi\right)
-a^D \delta m \bar \psi \psi
-\frac 12 a^{D-2} \delta Z_3
\eta^{\mu\nu}(\partial_\mu  \phi)(\partial_\nu \phi)
-\frac 12 a^D \delta m_\phi^2 \phi^2
\end{equation}
to the Lagrangian~(\ref{Lagrangian:deSitter}). Note that we do not discuss
vertex renormalisation here. These counterterms are de Sitter invariant,
provided $\delta Z_2$, $\delta m$, $\delta Z_3$ and $\delta m_\phi$ are
constant.
From ${\cal L}+\delta {\cal L}$, we can straightforwardly derive
Feynman rules~\cite{ProkopecWoodard:2003,GarbrechtProkopec:2006}
leading to the following self-energy function:
\begin{eqnarray}
\label{Sigma:bare}
-\ri \Sigma(x;x^\prime)&=&
\left(
-\ri f \mu^{\frac{\varepsilon}{2}}
a^{4-\varepsilon}
\right)
\ri S(x;x^\prime)
\left(
-\ri f \mu^{\frac{\varepsilon}{2}}
a^{4-\varepsilon}
\right)
\ri \Delta(x;x^\prime)
\\
&&
+
\ri \delta Z_2(aa^\prime)^{\frac{3-\varepsilon}2}
\ri \ps \delta^{4-\varepsilon}(x-x^\prime)
+\ri (aa^\prime)^{2-\frac{\varepsilon}2}
\delta m \delta^{4-\varepsilon}(x-x^\prime)
\nonumber
\\
&=&
-\frac{f^2 \mu^\varepsilon (aa^\prime)^{\frac 32}}{32\pi^{4-\varepsilon}}
\frac{
\Gamma\left(2- \frac \varepsilon2 \right)\Gamma\left(1-\frac \varepsilon2\right)}
{1-\frac \varepsilon2}
\ri \ps\frac 1{\left(\Delta x^2\right)^{2-\varepsilon}}
\nonumber\\
&&
-
\frac{f^2 \mu^\varepsilon (a a^\prime)^2}{16 \pi^{4-\varepsilon}}
\Gamma^2\left(1-\frac \varepsilon2\right)
\frac m{\left(\Delta x^2 \right)^{2-\varepsilon}}
\nonumber\\
&&+
\ri \delta Z_2(aa^\prime)^{\frac{3-\varepsilon}2}
\ri \ps \delta^{4-\varepsilon}(x-x^\prime)
- \ri(aa^\prime)^{2-\frac{\varepsilon}2}
\delta m \delta^{4-\varepsilon}(x-x^\prime)
\,.
\nonumber
\end{eqnarray}
In this expression, we have only kept terms which lead to ultraviolet
divergences and those which are used to renormalise them.
One may interpret it as the self-energy for a fermion of mass
$m$ coupled to a conformally coupled scalar ($m_\phi=0$ and $\xi=\frac 16$).
Important finite contributions arising for the minimally coupled case,
$\xi=0$, are
therefore not included here, but they are extensively discussed in
Ref.~\cite{ProkopecWoodard:2003} and in Ref.~\cite{GarbrechtProkopec:2006},
where a new mechanism for fermion mass generation is suggested.
As these contributions due to minimal coupling are also in the focus
of~\cite{ProkopecTornkvistWoodard:2002,ProkopecTornkvistWoodard:2002:2,ProkopecWoodard:2003:photon1,ProkopecWoodard:2002:photon2,ProkopecPuchwein:2004,ProkopecPuchwein:2005}, the main conclusions drawn in these previous articles remain
unaltered when applying the modifications we suggest in the following.

We now need to regulate terms which go as $1/\Delta x^4$ when
$\ve \rightarrow 0$, since they lead to logarithmic divergences when
integrated. Following Onemli and Woodard~\cite{OnemliWoodard:2002},
we make use of the $D$-dimensional representation of the Dirac
$\delta$-function
\begin{equation}
\label{delta:flat}
\partial^2 \frac 1{\Delta x_{++}^{D-2}}=\ri (D-2)\Omega_D \delta^D(\Delta x)
\,,
\end{equation}
where 
\begin{equation}
\Omega_D=\frac{2\pi^{D/2}}{\Gamma\left(\frac D2\right)}
\end{equation}
is the area of the $D$-dimensional unit sphere.
We then manipulate the logarithmic divergences as
\begin{eqnarray}
~\label{regulation}
\left(\frac{1}{\Delta x^2}\right)^{2-\varepsilon}
&=&
-\frac {1}{2\varepsilon(1-\varepsilon)}
\partial^2
\frac{1}{\Delta x^{2-2\varepsilon}}\\
&=&
-\frac{\tilde\partial^2}{2\varepsilon (1 - \varepsilon)}
\left[
\frac{1}{\Delta x^{2-2\varepsilon}}
-\frac {\varrho^\varepsilon}{\Delta x^{2-\varepsilon}}
\right]
-\varrho^\varepsilon \frac 1{2\varepsilon(1-\varepsilon)}
\partial^2\frac 1{\Delta x^{2-\varepsilon}}
\nonumber\\
&\approx&
-\tilde\partial^2\frac{\log\frac{\Delta x^2}{\varrho^2}}{4\Delta x^2}
-\ri \frac 1\varepsilon \left(1+\frac\varepsilon2\right)
\varrho^\varepsilon\frac{2\pi^{2-\frac\varepsilon2}}{\Gamma\left(2-\frac\varepsilon2\right)}
\delta^{4-\varepsilon}(\Delta x)
\nonumber
\,.
\end{eqnarray}
In the second step, we have added and subtracted the same contribution.
For dimensional reasons, similar to the point-splitting approach,
we have to introduce a regulator $\varrho$, which
has the the dimension of a length and may be a function of
the spacetime points $x_\mu$ and $x_\mu^\prime$.
The notation $\tilde\partial_\mu$ implies that this is a derivative which does
not act on the implicit dependence of $\varrho$ on $x_\mu$,
\begin{equation}
\tilde \partial_\mu \varrho(x;x^\prime)=0\,,
\end{equation}
and has been introduced to keep notation compact.

Since $\varrho$ corresponds to a \emph{comoving} length, we propose 
to employ a de Sitter-invariant constant \emph{physical}-length
regulator, which is given by
\begin{equation}
\label{Cov:Reg:dS}
\varrho(x;x^\prime)=\sqrt{\frac{1}{aa^\prime \mu^2}}\,,
\end{equation}
where $\mu$ is a constant regularisation scale.
The choice made in Ref.~\cite{OnemliWoodard:2002} is
\begin{equation}
\label{NonCov:Reg:dS}
\varrho=\mu^{-1}
\,,
\end{equation}
which corresponds to a physical length shrinking with the scale factor.
Of course, there is nothing wrong with this choice, because the
manipulation~(\ref{regulation})
just amounts to adding and subtracting the same contribution. Indeed,
tracing the calculation by Onemli and Woodard
further~\cite{OnemliWoodard:2002}, both terms  of the last expression
in~(\ref{regulation}) are fully taken into account. However, in
Refs.~\cite{ProkopecWoodard:2003,ProkopecTornkvistWoodard:2002,ProkopecTornkvistWoodard:2002:2,ProkopecWoodard:2003:photon1,ProkopecWoodard:2002:photon2,ProkopecPuchwein:2004,ProkopecPuchwein:2005,GarbrechtProkopec:2006}, the first
term is eventually neglected. This contribution however, with the choice
of the regulator~(\ref{NonCov:Reg:dS}), is not de Sitter invariant. Therefore,
a de Sitter breaking remainder in form of a local term occurs, which is
misinterpreted as anomalous~\cite{ProkopecWoodard:2003,ProkopecTornkvistWoodard:2002,ProkopecTornkvistWoodard:2002:2,ProkopecWoodard:2002:photon2,GarbrechtProkopec:2006}.

Applying the above procedure to the self-energy~(\ref{Sigma:bare}),
we find
\begin{eqnarray}
\label{Sigma:Ren}
\Sigma(x;x^\prime)&=&-\frac{f^2(aa^\prime)^{\frac 32}}{2^{7}\pi^4}
\tilde\ps \tilde\partial^2
\frac{\log \frac{\Delta x^2}{\varrho^2}}{\Delta x^2}
+
\ri \frac{f^2 (aa^\prime)^2}{2^6 \pi^4}
m\tilde\partial^2
\frac{\log \frac{\Delta x^2}{\varrho^2}}{\Delta x^2}
\\
&&
-\frac{f^2(aa^\prime)^{\frac 32} \mu^\ve \varrho^\varepsilon}{2^4\pi^{2-\frac\varepsilon2}}
\frac 1\varepsilon \frac{1+\frac\varepsilon2}{1-\frac\varepsilon2}
\Gamma\left(1-\frac\varepsilon2\right)\ri \ps \delta^{4-\varepsilon}(x-x^\prime)
\nonumber\\
&&
-\frac{f^2(aa^\prime)^{2} \mu^\ve \varrho^\varepsilon}{2^3\pi^{2-\frac\varepsilon2}}
\frac 1\varepsilon \frac{1+\frac\varepsilon2}{1-\frac\varepsilon2}
\Gamma\left(1-\frac\varepsilon2\right) m \delta^{4-\varepsilon}(x-x^\prime)
\nonumber\\
&&
- \delta Z_2(aa^\prime)^{\frac{3-\varepsilon}2}
\ri \ps \delta^{4-\varepsilon}(x-x^\prime)
- (aa^\prime)^{2-\frac{\varepsilon}2}
\delta m \delta^{4-\varepsilon}(x-x^\prime)
\,.
\nonumber
\end{eqnarray}
As a consistency check, we may convince ourselves that also for the momentum
space result in flat space, the infinite contributions
$\sim 1/\ve$ to the self energy
satisfy
\begin{equation}
\ve \frac{\partial \Sigma}{\partial q\!\!\!/}
=\frac 12 \ve \frac{\partial \Sigma}{\partial m} +O(\ve)\,.
\end{equation}

In order to render $\Sigma(x;x^\prime)$ finite, we choose the counterterms
\begin{eqnarray}
-\delta Z_2&=&\frac{1}{2^4 \pi^{2-\frac{\ve}{2}}} \frac{1}{\ve}
\frac{1+\frac{\ve}{2}}{1-\frac{\ve}{2}}
\Gamma\left(1-\frac{\ve}{2}\right)\,,
\\
\delta m &=&\frac{m}{2^3 \pi^{2-\frac{\ve}{2}}}
\frac{1}{\ve}\frac{1+\frac{\ve}{2}}{1-\frac{\ve}{2}}
\Gamma\left(1-\frac{\ve}{2}\right)\,.
\nonumber
\end{eqnarray}

With the regulator~(\ref{Cov:Reg:dS}), we obtain
\begin{eqnarray}
\label{Sigma:invariant}
\Sigma(x;x^\prime)&=&-\frac{f^2(aa^\prime)^{\frac 32}}{2^{7}\pi^4}
\tilde\ps \tilde\partial^2
\frac{\log \left(a a^\prime  \Delta x^2\right)}{\Delta x^2}
+
\ri \frac{f^2 (aa^\prime)^2}{2^6 \pi^4}
m\tilde\partial^2
\frac{\log \left( a a^\prime \Delta x^2\right)}{\Delta x^2}
\,.
\end{eqnarray}
$\Sigma(x;x^\prime)$ is de Sitter invariant if it scales as $(aa^\prime)^4$,
as can be seen explicitly
when employing it to  modify the conformally rescaled Dirac
equation~\cite{ProkopecWoodard:2003,GarbrechtProkopec:2006}.
Noting that $\Delta x^2$ scales as $(aa^\prime)^{-1}$
($y=aa^\prime \Delta x^2$ is de Sitter invariant) and that
furthermore the arguments of the logarithms are scale-invariant, this
property can indeed be verified for each of the two above contributions to
the self energy $\Sigma(x;x^\prime)$.

In contrast, when choosing the regulator~(\ref{NonCov:Reg:dS}), we find
\begin{eqnarray}
\label{Sigma:anomal}
\Sigma(x;x^\prime)&=&-\frac{f^2(aa^\prime)^{\frac 32}}{2^{7}\pi^4}
\tilde\ps \tilde\partial^2
\frac{\log \Delta x^2}{\Delta x^2}
+
\ri \frac{f^2 (aa^\prime)^2}{2^6 \pi^4}
m\tilde\partial^2
\frac{\log \Delta x^2}{\Delta x^2}
\\
&&
-\frac{f^2(aa^\prime)^{\frac 32}}{2^5 \pi^2}\log(aa^\prime)\ri \ps
\delta^4(x-x^\prime)
-\frac{f^2(aa^\prime)^2}{2^6 \pi^2}\log(aa^\prime) m
\delta^4(x-x^\prime)
\,.
\nonumber
\end{eqnarray}
None of the individual contributions are de Sitter invariant here, although
the sum is equal to the manifestly
invariant expression~(\ref{Sigma:invariant}). In the
work on Yukawa theory~\cite{ProkopecWoodard:2003,GarbrechtProkopec:2006},
the first two terms of~(\ref{Sigma:anomal}) are not discussed further and
it is incorrectly claimed that the third and fourth term break de Sitter
invariance anomalously. The same applies to the corresponding terms
being identified as anomalous in the work on scalar quantum
electrodynamics~\cite{ProkopecTornkvistWoodard:2002,ProkopecTornkvistWoodard:2002:2,ProkopecWoodard:2002:photon2}.
We emphasise however that the putatively anomalous terms are not in the main
focus of the papers~\cite{ProkopecWoodard:2003,ProkopecTornkvistWoodard:2002,ProkopecTornkvistWoodard:2002:2,ProkopecWoodard:2002:photon2,GarbrechtProkopec:2006}, but are rather treated as a side effect. What we point out here is that
the ultraviolet regularisation can be taken out without introducing additional
de Sitter breaking contributions.

It would of course be desirable to obtain a manifestly covariant expression
for $\Sigma(x;x^\prime)$ in general backgrounds,
as may be constructed employing DeWitt's
technique for expanding two-point functions in terms of the geodesic
distance~\cite{DeWitt:1975}.
Future work will show whether this leads to a fairly simple and manageable
result.

An immediate consequence of the correct application of the ultraviolet
regularisation procedure is of course the vanishing of de Sitter breaking terms
of ultraviolet origin found for Yukawa theory in
Refs.~\cite{ProkopecWoodard:2003,GarbrechtProkopec:2006}, which have been
misinterpreted as anomalous\footnote{
Explicitly, the second term in Eqn.~(14) of
Ref.~\cite{GarbrechtProkopec:2006} can be combined with the first term in
a de Sitter invariant way. The same statement applies
for the second term of Eqn.~(24) of Ref.~\cite{ProkopecWoodard:2003}, where
Yukawa interactions of a minimally coupled massless scalar field are
investigated.
Note that in the latter case, the additional de Sitter breaking
contribution originating from the scalar propagator persists even when taking
account of the absence of an anomoaly.
}.
However, Onemli's and Woodard's regularisation procedure is directly
applicable to all calculations involving logarithmic divergences.
For example,
in the original work~\cite{OnemliWoodard:2002}, Onemli and Woodard
study $\phi^4$-theory. Note however, that they use a de Sitter
breaking scalar propagator, as necessary for a massless scalar
field~\cite{AllenFolacci:1987}, which reads in $D=4$ space-time
dimensions~\cite{OnemliWoodard:2002}
\begin{equation}
\label{MinMlessProp}
\ri \Delta(x;x^\prime)=\frac{H^2}{4\pi^2}\left\{
\frac{1}{y(x;x^\prime)}-\frac 12 \log y(x;x^\prime)
+ \frac 12 \log (aa^\prime)
\right\}
\,.
\end{equation}
The last term obviously breaks de Sitter
invariance,
but is of an utterly different, namely infrared, origin. In loop calculations
for renormalisable theories such as Yukawa or $\phi^4$, it does not lead
to ultraviolet divergences, but yet gives rise to important finite de Sitter
breaking effects. Note however
that the focus of the present paper is not on the discussion of the
peculiarities of the massless minimally coupled scalar field
but on the regularisation of logarithmic
ultraviolet divergences resulting from the leading term $\propto 1/y$
of the scalar propagator, which is universal for scalar fields of different
mass and curvature couplings.

A detailed and excellent discussion of these different
-- ultraviolet and infrared ---
de Sitter breaking contributions for the example of
scalar electrodynamics can be found in
Ref.~\cite{ProkopecTornkvistWoodard:2002:2}, where however the conformal
anomaly of gauge theory is inappropriately made responsible for the ultraviolet
induced breakdown of de Sitter invariance. We point out that employing the
invariant regularisation as procedure proposed here,
ultraviolet induced de Sitter breaking
can also be casted off from scalar electrodynamics~\cite{ProkopecTornkvistWoodard:2002,ProkopecTornkvistWoodard:2002:2,ProkopecWoodard:2003:photon1,ProkopecWoodard:2002:photon2,ProkopecPuchwein:2004,ProkopecPuchwein:2005}.

Let us finally treat the wave function renormalisation of the scalar field,
since this is the case which is important for the generation of density
perturbations during inflation.
In order to do so, it turns out that we need to keep track of the
dependence of $\ri S_\pm(x;x^\prime)$ on $\ve\not=0$ up to logarithmic
order ({\it cf.} Eqn.~(\ref{Spm:Expanded})),
\begin{eqnarray}
\ri S_\pm(x;x^\prime)&=&
\frac{H^{2-\varepsilon}}{4\pi^{2 -\frac{\ve}{2}}}
\Gamma\left(1-\frac\ve2\right)\frac{1}{y^{1-\frac{\ve}{2}}}
\\
&&
+
\frac{H^{-\varepsilon}}{16\pi^{2 -\frac{\ve}{2}}}
\Gamma\left(1-\frac\ve2 \right)
\left(\mp\ri H m + m^2\right)
\nonumber\\
&&
\times
\left\{
\frac 2\ve ( y^\frac\ve2 -1 )
-2\log 2 +\psi\left(1 \mp \ri \frac mH\right)
+\psi\left(2 \pm \ri \frac mH\right)
-1 + 2\gamma_{\rm E}
\right\}
\,.
\nonumber
\end{eqnarray}
Then, the divergent contributions to the scalar self energy
$\ri \Pi(x;x^\prime)$ can be regularised as (\emph{cf.} the fermionic
case, Eqn.~(\ref{Sigma:Ren}))
\begin{eqnarray}
\label{Pi:Ren}
\Pi(x;x^\prime)
&=&\ri \frac{f^2 aa^\prime}{2^7 \pi^4}\tilde \partial^4 
\frac{\log \frac{\Delta x^2}{\varrho^2}}{\Delta x^2}
+
\ri \frac{f^2 (aa^\prime)^2}{2^5 \pi^4} m^2 \tilde \partial^2 
\frac{\log \frac{\Delta x^2}{\varrho^2}}{\Delta x^2}
\\
&&
-\frac{f^2 aa^\prime \mu^\ve \varrho^\varepsilon}{2^4\pi^{2-\frac\varepsilon2}}
\frac 1\varepsilon \frac{1+\frac\varepsilon2}{1-\frac\varepsilon2}
\Gamma\left(1-\frac\varepsilon2\right) \partial^2 \delta^{4-\varepsilon}(x-x^\prime)
\nonumber\\
&&
-\frac{f^2 (aa^\prime)^2 \mu^\ve \varrho^\varepsilon}{2^2\pi^{2-\frac\varepsilon2}}
\frac 1\varepsilon \frac{1+\frac\varepsilon2}{1-\frac\varepsilon2}
\Gamma\left(1-\frac\varepsilon2\right) m^2 \delta^{4-\varepsilon}(x-x^\prime)
\nonumber
\nonumber\\
&&
+\delta Z_3 (aa^\prime)^\frac{2-\ve}{2} \partial^2 \delta^{4-\ve}(x-x^\prime)
+(aa^\prime)^\frac{4 -\ve}{2} \delta m_\phi^2 \delta^{4-\ve}(x-x^\prime)
\,,
\end{eqnarray}
which is renormalised by the counterterms
\begin{eqnarray}
\delta Z_3&=&\frac{1}{2^4 \pi^{2-\frac{\ve}{2}}} \frac{1}{\ve}
\frac{1+\frac{\ve}{2}}{1-\frac{\ve}{2}}
\Gamma\left(1-\frac{\ve}{2}\right)\,,
\\
\delta m_\phi^2 &=&\frac{m_\phi^2}{2^2 \pi^{2-\frac{\ve}{2}}}
\frac{1}{\ve}\frac{1+\frac{\ve}{2}}{1-\frac{\ve}{2}}
\Gamma\left(1-\frac{\ve}{2}\right)\,.
\nonumber
\end{eqnarray}
Again, a de Sitter invariant expression is obtained when using the
regulator~(\ref{Cov:Reg:dS}) and consistency with the flat space result
derived in momentum space
\begin{equation}
\ve \frac{\partial \Pi}{\partial q^2}
=\frac 14 \ve \frac{\partial \Pi}{\partial m^2} + O(\ve)
\end{equation}
is found.

Of course, we would like to reproduce now the contribution
to $V_{\rm eff}^\psi$ from $\Pi(x;x^\prime)$. At first glance, it is disturbing
that there appear to be no divergences in $\Pi(x;x^\prime)$ which are
proportional to $m^2H^2$, while these terms occur in
$V_{\rm eff}^\psi$, Eqn.~(\ref{Veff:fermion}).
However, when calculating the potential from $\Pi(x;x^\prime)$ and
keeping only the divergences $\sim 1/\ve$, we observe
when setting $m=f\phi$
\begin{eqnarray}
V_\Pi^\psi&=&\frac 12 \frac 1{a^4}\int d^4 x^\prime f\phi(x) \phi(x;x^\prime)
f\phi(x^\prime)+O(\ve^0)
\\
&=&
- \frac 1\ve \frac 1{a^4}\int d^4 x^\prime
\left\{
\frac 1{8\pi^2} m^4 (aa^\prime)^2 \delta^4(x-x^\prime)
+\frac 1{32\pi^2} m^2 aa^\prime \partial^2 \delta^4(x-x^\prime)
\right\}+O(\ve^0)
\nonumber
\\
&=&
-\frac 1\ve \left(\frac 1{8\pi^2} m^4 + \frac 1{16\pi^2} m^2 H^2\right)
\nonumber+O(\ve^0)
\,.
\end{eqnarray}
Hence we have recovered a divergence $\sim m^2H^2$. Note however that a
full reconstruction of the effective potential at leading order
will also involve tadpole diagrams. Furthermore, we completely miss the
terms $\propto \log m$.
We suspect that the solution to this problem lies in the treatment
of the terms in the first line of Eqn.~(\ref{Pi:Ren}) and equivalently
of Eqn.~(\ref{Sigma:Ren}), which correspond to expressions
$\propto \log(F(m,q))$,
which are familiar from momentum space and where $F(m,q)$ is some function.

\section{Conclusions}

In this paper, we rederive the Candelas-Raine effective
potentials~\cite{CandelasRaine:1975} for fermions
and for scalars. The expressions we find are benign as $H\rightarrow 0$
and moreover they reduce to the Coleman-Weinberg form in that limit. The
different curvature coupling of fermions and scalars gives rise to
curvature-induced supersymmetry breaking, evident from the potential
sum~(\ref{Veff:SUSY}). This may be of relevance for inflationary model
building.

We also point out that the ultraviolet regularisation procedure by
Onemli and Woodard~\cite{OnemliWoodard:2002} does not lead to an anomalous
breakdown of de Sitter invariance.
Yet, we have to resolve the question how the remaining
nonanalytic terms $\propto \log \Delta x^2$ have to be evaluated.
Techniques to approach this problem have been developed and successfully
applied to compute the effect of self-energy corrections on the
equations of motion of various quantum
fields~\cite{ProkopecTornkvistWoodard:2002,ProkopecTornkvistWoodard:2002:2,ProkopecWoodard:2003:photon1,ProkopecWoodard:2002:photon2,ProkopecPuchwein:2004,ProkopecPuchwein:2005,ProkopecWoodard:2003,GarbrechtProkopec:2006}.
We emphasise, that position space techniques are not only useful for
computations of effects within curved space-times, but for any kind
of backgrounds, for example inhomogeneous electric
fields~\cite{FriedWoodard:2001,GiesKlingmuller:2005}.

The possibility of an invariant renormalisation implies that within
field theory, a transplanckian problem does not necessarily exist,
because all processes can be imposed to be fully de Sitter invariant
by the choice of an invariant renormalisation scheme. This is a consequence
of the manifest coordinate invariance of the underlying
Lagrangian~(\ref{Lagrangian}) and that this invariance does not appear
to be broken by perturbation theory anomalies.
Since there is no observational evidence
for a breaking of Lorentz symmetry in flat space, where one therefore
routinely uses covariant regularisation procedures,
one may consider also in curved space-times covariant regularisation as a
natural choice.  Note however that within string theory, albeit also being
formulated in a manifestly covariant way, a breakdown of de Sitter
invariance
leading to signatures in the primordial perturbation spectrum
is expected~\cite{Kempf:2000,EastherGreeneKinneyShiu:2001,EastherGreeneKinneyShiu:2001:2}. A discussion of the renormalisation of these boundary effects
and some aspects of renormalisation of self-interacting scalar theories,
which may be compared to the results presented here, is
provided in Refs.~\cite{CollinsHolman:2005,CollinsHolman:2005:2}.

We emphasise
that the ultraviolet regularisation in de Sitter space is not directly
related to the breakdown of de Sitter symmetry for massless minimally coupled
scalar fields~\cite{AllenFolacci:1987,TsamisWoodard:1993}.
It is a tree level effect and is due to the fact that it is not
possible to construct a de Sitter invariant Green function which takes the
Hadamard form in this case.
It corresponds to an infrared divergence, which becomes manifest
when taking $m_\phi \rightarrow 0$ and $\xi \rightarrow 0$ in
Eqns.~(\ref{Prop:Scalar}) and ~(\ref{nuD}).
Because of its infrared origin, this type of
de Sitter breaking for massless minimally coupled fields is not related
to the transplanckian problem.

Concerning the prospects of the work presented here,
a goal of future efforts should be to obtain a dictionary translating from
position space expressions to their familiar momentum space counterparts,
such that new effects in curved space-time may reliably be identified.
Furthermore,
a covariant generalisation of the de Sitter space results to more general
backgrounds is desirable. While being interesting on their own behalf,
quantum loop effects in curved space-time
may give relevant input to inflationary model building or even directly lead to
observational consequences, such that routine techniques for their
calculation are of great importance.


\begin{thebibliography}{999}

\bibitem{ChibisovMukhanov:1981}
  V.~F.~Mukhanov and G.~V.~Chibisov,
  ``Quantum Fluctuation And 'Nonsingular' Universe. (In Russian),''
  JETP Lett.\  {\bf 33} (1981) 532
  [Pisma Zh.\ Eksp.\ Teor.\ Fiz.\  {\bf 33} (1981) 549].

\bibitem{Starobinsky:1982}
  A.~A.~Starobinsky,
  ``Dynamics Of Phase Transition In The New Inflationary Universe Scenario And
  Generation Of Perturbations,''
  Phys.\ Lett.\ B {\bf 117} (1982) 175.

\bibitem{Hawking:1982}
  S.~W.~Hawking,
  ``The Development Of Irregularities In A Single Bubble Inflationary
  Universe,''
  Phys.\ Lett.\ B {\bf 115} (1982) 295.

\bibitem{GuthPi:1982}
  A.~H.~Guth and S.~Y.~Pi,
  ``Fluctuations In The New Inflationary Universe,''
  Phys.\ Rev.\ Lett.\  {\bf 49} (1982) 1110.

\bibitem{BardeenSteinhardtTurner:1983}
  J.~M.~Bardeen, P.~J.~Steinhardt and M.~S.~Turner,
  ``Spontaneous Creation Of Almost Scale - Free Density Perturbations In An
  Inflationary Universe,''
  Phys.\ Rev.\ D {\bf 28} (1983) 679.

\bibitem{Spergel:2006}
  D.~N.~Spergel {\it et al.},
  ``Wilkinson Microwave Anisotropy Probe (WMAP) three year results:
  Implications for cosmology,''
  arXiv:astro-ph/0603449.

\bibitem{ColemanWeinberg:1973}
  S.~R.~Coleman and E.~Weinberg,
  ``Radiative Corrections As The Origin Of Spontaneous Symmetry Breaking,''
  Phys.\ Rev.\ D {\bf 7} (1973) 1888.

\bibitem{DeWitt:1975}
  B.~S.~DeWitt,
  ``Quantum Field Theory In Curved Space-Time,''
  Phys.\ Rept.\  {\bf 19} (1975) 295.

\bibitem{CandelasRaine:1975}
  P.~Candelas and D.~J.~Raine,
  ``General Relativistic Quantum Field Theory - An Exactly Soluble Model,''
  Phys.\ Rev.\ D {\bf 12} (1975) 965.

\bibitem{MiaoWoodard:2006}
  S.~P.~Miao and R.~P.~Woodard,
  ``Leading log solution for inflationary Yukawa,''
  arXiv:gr-qc/0602110.

\bibitem{Prokopec:2006}
  T.~Prokopec,
  ``A solution to the cosmological constant problem,''
  arXiv:gr-qc/0603088.



\bibitem{ProkopecWoodard:2003}
  T.~Prokopec and R.~P.~Woodard,
  ``Production of massless fermions during inflation,''
  JHEP {\bf 0310} (2003) 059
  [arXiv:astro-ph/0309593];
  ibid. errata.

\bibitem{ProkopecTornkvistWoodard:2002}
  T.~Prokopec, O.~Tornkvist and R.~P.~Woodard,
  ``Photon mass from inflation,''
  Phys.\ Rev.\ Lett.\  {\bf 89} (2002) 101301
  [arXiv:astro-ph/0205331].

\bibitem{ProkopecTornkvistWoodard:2002:2}
  T.~Prokopec, O.~Tornkvist and R.~P.~Woodard,
  ``One loop vacuum polarization in a locally de Sitter background,''
  Annals Phys.\  {\bf 303} (2003) 251
  [arXiv:gr-qc/0205130].

\bibitem{ProkopecWoodard:2003:photon1}
  T.~Prokopec and R.~P.~Woodard,
  ``Vacuum polarization and photon mass in inflation,''
  Am.\ J.\ Phys.\  {\bf 72} (2004) 60
  [arXiv:astro-ph/0303358].

\bibitem{ProkopecWoodard:2002:photon2}
  T.~Prokopec and R.~P.~Woodard,
  ``Dynamics of super-horizon photons during inflation with vacuum
  polarization,''
  Annals Phys.\  {\bf 312} (2004) 1
  [arXiv:gr-qc/0310056].

\bibitem{ProkopecPuchwein:2004}
  T.~Prokopec and E.~Puchwein,
  ``Nearly minimal magnetogenesis,''
  Phys.\ Rev.\ D {\bf 70} (2004) 043004
  [arXiv:astro-ph/0403335].

\bibitem{ProkopecPuchwein:2005}
  T.~Prokopec and E.~Puchwein,
  ``Photon mass generation during inflation: de Sitter invariant case,''
  JCAP {\bf 0404} (2004) 007
  [arXiv:astro-ph/0312274].

\bibitem{GarbrechtProkopec:2006}
  B.~Garbrecht and T.~Prokopec,
  ``Fermion mass generation in de Sitter space,''
  Phys.\ Rev.\ D {\bf 73} (2006) 064036
  [arXiv:gr-qc/0602011].

\bibitem{OnemliWoodard:2002}
  V.~K.~Onemli and R.~P.~Woodard,
  ``Super-acceleration from massless, minimally coupled $\phi^4$,''
  Class.\ Quant.\ Grav.\  {\bf 19} (2002) 4607
  [arXiv:gr-qc/0204065].


\bibitem{BrandenbergerMartin:2000}
  R.~H.~Brandenberger and J.~Martin,
  ``The robustness of inflation to changes in super-Planck-scale physics,''
  Mod.\ Phys.\ Lett.\ A {\bf 16} (2001) 999
  [arXiv:astro-ph/0005432].

\bibitem{BrandenbergerMartin:2000:2}
  J.~Martin and R.~H.~Brandenberger,
  ``The trans-Planckian problem of inflationary cosmology,''
  Phys.\ Rev.\ D {\bf 63} (2001) 123501
  [arXiv:hep-th/0005209].

\bibitem{Kempf:2000}
  A.~Kempf,
  ``Mode generating mechanism in inflation with cutoff,''
  Phys.\ Rev.\ D {\bf 63} (2001) 083514
  [arXiv:astro-ph/0009209].

\bibitem{CollinsHolman:2005}
  H.~Collins and R.~Holman,
  ``Renormalization of initial conditions and the trans-Planckian problem  of
  inflation,''
  Phys.\ Rev.\ D {\bf 71} (2005) 085009
  [arXiv:hep-th/0501158].

\bibitem{others}
  This result has most probably appeared earlier in the literature.

\bibitem{CollinsHolman:2005:2}
  H.~Collins and R.~Holman,
  ``An effective theory of initial conditions in inflation,''
  arXiv:hep-th/0507081.

\bibitem{DaviesFullingUnruh:1976}
  P.~C.~W.~Davies, S.~A.~Fulling and W.~G.~Unruh,
  ``Energy Momentum Tensor Near An Evaporating Black Hole,''
  Phys.\ Rev.\ D {\bf 13} (1976) 2720.

\bibitem{Christensen:1976}
  S.~M.~Christensen,
  ``Vacuum Expectation Value Of The Stress Tensor In An Arbitrary Curved
  Background: The Covariant Point Separation Method,''
  Phys.\ Rev.\ D {\bf 14} (1976) 2490.

\bibitem{Christensen:1978}
  S.~M.~Christensen,
  ``Regularization, Renormalization, And Covariant Geodesic Point Separation,''
  Phys.\ Rev.\ D {\bf 17} (1978) 946.

\bibitem{ElizaldeOdintsov:1995}
  E.~Elizalde and S.~D.~Odintsov,
  ``The Higgs-Yukawa model in curved space-time,''
  Phys.\ Rev.\ D {\bf 51} (1995) 5950
  [arXiv:hep-th/9503111].

\bibitem{ChernikovTagirov:1968}
  N.~A.~Chernikov and E.~A.~Tagirov,
  ``Quantum Theory Of Scalar Fields In De Sitter Space-Time,''
  Annales Poincare Phys.\ Theor.\ A {\bf 9} (1968) 109.

\bibitem{CopelandLiddleLythStewartWands:1994}
  E.~J.~Copeland, A.~R.~Liddle, D.~H.~Lyth, E.~D.~Stewart and D.~Wands,
  ``False vacuum inflation with Einstein gravity,''
  Phys.\ Rev.\ D {\bf 49} (1994) 6410
  [arXiv:astro-ph/9401011].

\bibitem{DineRandallThomas:1995}
  M.~Dine, L.~Randall and S.~D.~Thomas,
  ``Baryogenesis from flat directions of the supersymmetric standard model,''
  Nucl.\ Phys.\ B {\bf 458} (1996) 291
  [arXiv:hep-ph/9507453].

\bibitem{Panagiotakopoulos:1997}
  C.~Panagiotakopoulos,
  ``Blue perturbation spectra from hybrid inflation with canonical
  supergravity,''
  Phys.\ Rev.\ D {\bf 55} (1997) 7335
  [arXiv:hep-ph/9702433].

\bibitem{Panagiotakopoulos:2004}
  C.~Panagiotakopoulos,
  ``Realizations of hybrid inflation in supergravity with natural initial
  conditions,''
  Phys.\ Rev.\ D {\bf 71} (2005) 063516
  [arXiv:hep-ph/0411143].

\bibitem{JeannerotPostma:2005}
  R.~Jeannerot and M.~Postma,
  ``Confronting hybrid inflation in supergravity with CMB data,''
  JHEP {\bf 0505} (2005) 071
  [arXiv:hep-ph/0503146].

\bibitem{BasteroGilKingShafi:2006}
  M.~Bastero-Gil, S.~F.~King and Q.~Shafi,
  arXiv:hep-ph/0604198.

\bibitem{KoldaMarch-Russell:1998}
  C.~F.~Kolda and J.~March-Russell,
  ``Supersymmetric D-term inflation, reheating and Affleck-Dine
  baryogenesis,''
  Phys.\ Rev.\ D {\bf 60} (1999) 023504
  [arXiv:hep-ph/9802358].

\bibitem{BuchbinderOdintsov:1985}
For a study of nonsupersymmetric gauge theory in general backgrounds, see:
  I.~L.~Buchbinder and S.~D.~Odintsov,
  ``Effective Potential And Phase Transitions Induced By Curvature In Gauge
  Theories In Curved Space-Time,''
  Yad.\ Fiz.\  {\bf 42} (1985) 1268
  [Class.\ Quant.\ Grav.\  {\bf 2} (1985) 721].
These results should be carefully reevaluated before applying them to
cosmology, since they do not reproduce the Coleman-Weinberg potentials for
zero curvature.

\bibitem{AllenFolacci:1987}
  B.~Allen and A.~Folacci,
  ``The Massless Minimally Coupled Scalar Field In De Sitter Space,''
  Phys.\ Rev.\ D {\bf 35} (1987) 3771.

\bibitem{FriedWoodard:2001}
  H.~M.~Fried and R.~P.~Woodard,
  ``The one loop effective action of QED for a general class of electric
  fields,''
  Phys.\ Lett.\ B {\bf 524} (2002) 233
  [arXiv:hep-th/0110180].

\bibitem{GiesKlingmuller:2005}  
  H.~Gies and K.~Klingm\"uller,
  ``Pair production in inhomogeneous fields,''
  Phys.\ Rev.\ D {\bf 72} (2005) 065001
  [arXiv:hep-ph/0505099].

\bibitem{EastherGreeneKinneyShiu:2001}
  R.~Easther, B.~R.~Greene, W.~H.~Kinney and G.~Shiu,
  ``Inflation as a probe of short distance physics,''
  Phys.\ Rev.\ D {\bf 64} (2001) 103502
  [arXiv:hep-th/0104102].

\bibitem{EastherGreeneKinneyShiu:2001:2}
  R.~Easther, B.~R.~Greene, W.~H.~Kinney and G.~Shiu,
  ``Imprints of short distance physics on inflationary cosmology,''
  Phys.\ Rev.\ D {\bf 67} (2003) 063508
  [arXiv:hep-th/0110226].

\bibitem{TsamisWoodard:1993}
  N.~C.~Tsamis and R.~P.~Woodard,
  ``The Physical basis for infrared divergences in inflationary quantum
  gravity,''
  Class.\ Quant.\ Grav.\  {\bf 11} (1994) 2969.

\end{thebibliography}
\end{document}